\documentclass[aps,showpacs,nofootinbib]{revtex4}
\usepackage{amsmath} 
\usepackage{amssymb}
\usepackage{graphicx} 
\usepackage{color} 
\usepackage{epstopdf}

\begin{document}

\newcommand{\blue}[1]{\textcolor{blue}{#1}}
\newcommand{\new}{\blue}
\newcommand{\green}[1]{\textcolor{green}{#1}}
\newcommand{\modif}{\green}
\newcommand{\red}[1]{\textcolor{red}{#1}}
\newcommand{\attention}{\red}


\title{Single top quark production at ultra-high energies}

\author{V. A. Okorokov} \email{VAOkorokov@mephi.ru; Okorokov@bnl.gov}
\affiliation{National Research Nuclear University MEPhI (Moscow
Engineering Physics Institute), Kashirskoe highway 31, 115409
Moscow, Russia}

\date{\today}

\begin{abstract}
The processes with single top quark production provide a prototype
search for the types of final state that are expected in many new
physics scenarios. Some distinctive features are considered for
particle production in the top sector in ultra-high energy domain
which can be covered partly in the collisions of cosmic ray
particles with atmosphere. The single top quark production through
electroweak interaction is studied within the Standard Model and
the Effective Field Theory approach used for calculations of total
partonic cross sections in $s$--channel. These quantitative
results can be important for both the future collider experiments
at center-of-mass energy frontier and the improvement of the
phenomenological models for development of the cosmic ray cascades
in ultra-high energy domain. Thus the study allows the better
understanding of heavy particle production and emphasizes the
exciting interrelation between the high-energy physics on
accelerators and ultra-high energy cosmic ray measurements.
\end{abstract}


\maketitle

\section{Introduction}\label{sec:0}

Among the most challenging problems for the modern physics of
fundamental interactions is search for the physics beyond the
Standard Model (SM) and the study of the deconfined quark--gluon
matter under extreme conditions called also quark--gluon plasma
(QGP) which can be created in subatomic particle collisions at
high enough energies. The top quark ($t$) sector plays an
important role in searches for new physics due to largest mass of
$t$ among fundamental particles of the SM and consequent its
enhanced sensitivity to hypothetical new heavy particles and
interactions. At present a particles beyond the SM are not
observed at the Large Hadron Collider (LHC) and this situation
implies that there is a considerable energy gap between the SM
particles and new physics. Due to the scale separation, various
physics effects beyond SM (BSM) below the energy threshold of new
physics particle can be characterized by the model-independent
Effective Field Theory (EFT) framework. The study of top is
crucially important for future development of EFT approach and
constraint of its parameters. The uniqueness of the top quark is
not only due to its heavy mass, but also due to the fact that it
is the only quark that decays before it can hadronize. Thus the
study of $t$ behavior in hot environment created at ultra-high
energies opens a new ways for investigation of, in particular, a
very early pre-equilibrium stages of space--time evolution of QGP.
Measurements of interactions of ultra-high energy cosmic rays
(UHECR), i.e. cosmic ray particles with initial laboratory
energies larger than 0.1--1 EeV, with nuclei in the atmosphere
allow the new unique possibilities for study of multiparticle
production processes at energies (well) above not only the LHC
range but future collider on Earth as well. Due to the air
composition and main components of the UHECR the passage of UHECR
particles through atmosphere can be considered as collision mostly
small systems. It should be emphasized collisions at ultra-high
energies can lead to creation of QGP even in light nuclear
interactions \cite{Okorokov-PAN-82-838-2019}. Therefore the study
of single $t$ production at ultra-high energies seems important
for search for the signatures of physics BSM and possible creation
of bubble of QGP in small system collisions.

\section{Formalism for single top production}\label{sec:1}

Signle top quark is produced through electroweak (EW)
interactions. The total inclusive single $t$ production cross
section in (anti)proton--proton ($pp$, $\bar{p}p$) collision can
be written, in particular, as follows
\begin{equation}
\sigma_{\scriptsize{\mbox{tot}}}^{t}=\sum_{i,j}\int_{0}^{1}dx_{1}dx_{2}\,f_{i}(x_{1},\mu_{F})
f_{j}(x_{2},\mu_{F})\hat{\sigma}_{ij}(m_{W},m_{t},s,\mu_{F}^{2},\mu_{R}^{2})\delta(s-Q^{2}),
\label{eq:1.2}
\end{equation}
Here $\mu_{F,R}$ are the factorization and renormalization scales,
$i,j$ run over all initial state partons contributed in the
production channel under discussion, $\forall\,k = 1, 2: x_{k}$ is
the fraction of the 4-momentum of incoming hadron carried out by
the parton, $f_{i}(x_{k},\mu_{F})$ is the distribution function
for (anti)parton $i$, $m_{W}$, $m_{t}$ are the mass of $W$ boson
and $t$ quark, $s=x_{1}x_{2}s_{\scriptsize{p(\bar{p})p}}$ is
squared partonic c.m. energy with $s_{\scriptsize{p(\bar{p})p}}$
is the square of the c.m. energy of the colliding particles,
namely $p(\bar{p})$ here, $Q$ is the mass of virtual boson ($W$).
The (anti)parton distribution functions (PDFs) are multiplied on
the total partonic (short-distance) cross section
$\hat{\sigma}_{ij}$ for the single $t$ production from partons
$i,j$. Here following choice is used for the factorization and
renormalization scales $\mu_{F}=\mu_{R}\equiv \mu$, $\mu=m_{t}$;
and, as previously \cite{Okorokov-JPCS-1690-012006-2020}, the
fixed value $x_{1}x_{2}=1/9$ is chosen in order to get the
well-known relation between $e^{+}e^{-}$ and partonic process
$s_{\scriptsize{e^{+}e^{-}}}=s$. One can note the information
about PDFs is very limited and model--dependent at ultra-high
energies $\sqrt{\smash[b]{s_{\scriptsize{pp}}}} \geq 0.1$ PeV.
That amplifies the uncertainties for hadronic cross section
(\ref{eq:1.2}) significantly. Thus the partonic cross section
$\hat{\sigma}_{ij}(m_{W},m_{t},s)$ is in the main quantity for
study in the present work.

Taking into account the relative strengths of the EW interactions
with quark mixing the following partonic subprocesses mostly
contribute in the single $t$ production \cite{PRD-74-114012-2006}
\begin{subequations}
\begin{equation}
u+\bar{d} \to t+\bar{b}, \label{eq:1.1.1}
\end{equation}
\begin{equation}
u+b \to t+d, \label{eq:1.1.2a}
\end{equation}
\begin{equation}
\bar{d}+b \to t+\bar{u}, \label{eq:1.1.2b}
\end{equation}
\begin{equation}
b+g \to t+W. \label{eq:1.1.3}
\end{equation}\label{eq:1.1}
\end{subequations}
\hspace{-0.4cm} The partonic collision (\ref{eq:1.1.1}) is the
$s$--channel process considered in the present work. The
subprocesses (\ref{eq:1.1.2a}) and (\ref{eq:1.1.2b}) correspond to
the $t$--channel and type of partonic interaction (\ref{eq:1.1.3})
is the $tW$ associated production.

Within the approach of vanishing all quark masses except $m_{t}$
($\forall\,f \geq b: m_{f} \to 0$, $f$ is the quark flavor) the
dominant contribution to the leading order (LO) partonic cross
section for single $t$ production in the $s$--channel in SM is
described by the formula
\cite{PRD-83-034006-2011,JHEP-2002-131-2020}
\begin{equation}
\displaystyle \hat{\sigma}^{\scriptsize{\mbox{(0),EW}}}_{u\bar{d}
\to t\bar{b}}(m_{W},m_{t},s)=\Pi_{V}^{2}\frac{\pi
\alpha_{W}^{2}}{24s}\frac{\beta_{ts}^{4}(2+\rho_{ts})}{\beta_{Ws}^{4}},
\label{eq:1.3}
\end{equation}
where $\forall\,x=W, t: \rho_{xs}=m_{x}^{2}/s$,
$\beta_{xs}^{2}=1-\rho_{xs}$ in accordance with the common
notation basis with top pair production
\cite{Okorokov-JPCS-1690-012006-2020}; $\alpha_{W}(\mu)$ is the
SU(2) running coupling associated with $q\bar{q}W$ vertex and the
constant $\alpha_{W}(\mu)$ is renormalized with $N_{f}=5$ active
flavors; $\Pi_{V} \equiv |V_{ud}||V_{tb}|$ is the product of the
CKM elements for $u\bar{d}$ and $t\bar{b}$ vertices.

According to the detailed discussion elsewhere
\cite{Okorokov-JPCS-1690-012006-2020}, effects BSM can be
described within EFT approach with the general form of Lagrangian
$\mathcal{L}_{\scriptsize{\mbox{EFT}}}=\sum_{j=0}\mathcal{L}_{j}\Lambda^{-j}$,
where $\mathcal{L}_{0}$ is the SM Lagrangian and
$\mathcal{L}_{\scriptsize{\mbox{eff}}}=\sum_{j=1}\mathcal{L}_{j}\Lambda^{-j}$
-- effective part containing the effects of new physics, $\Lambda$
is the energy scale of the possible physics BSM. The leading
contributions arise at dimension six\footnote{Contributions from
odd--dimensional operators lead to lepton- and baryon-number
violation \cite{AP-335-21-2013} and are neglected in this work. In
particular, operator of dimension five, called also Weinberg
operator, can be interested for neutrino physics but it is
insignificant for the energy range considered within the present
work below \cite{Okorokov-JPCS-1690-012006-2020,PU-65-653-2022}.}
and can be parameterized in terms of Wilson coefficients
$C_{k}^{(6)}$ of dimension--6 operators $O_{k}^{(6)}$ in the
effective part
$\mathcal{L}_{\scriptsize{\mbox{eff}}}=\mathcal{L}_{\scriptsize{\mbox{eff}}}^{(0)}+\mathcal{O}(\Lambda^{-4})$,
$\mathcal{L}_{\scriptsize{\mbox{eff}}}^{(0)}=\sum_{k}\bigl(C_{k}^{(6)}\Lambda^{-2}{}^{+}O_{k}^{(6)}+
\mbox{h.c.}\bigr)+\sum_{l}C_{l}^{(6)}\Lambda^{-2}O_{l}^{(6)}$,
where the sum runs over all operators corresponding to the
interaction processes under consideration and non-hermitian
operators are denoted as ${}^{+}O$ \cite{JHEP-2002-131-2020}. The
partonic subprocesses considered here involve $t$-quark and the
lists of the dimension--6 operators for $t$ production can be
found elsewhere \cite{PRD-83-034006-2011,JHEP-2002-131-2020}. In
general the Wilson coefficients are free parameters by definition
and are constrained by experimental measurements. Truncation of
$\mathcal{L}_{\scriptsize{\mbox{eff}}}$ by only leading
contributions, i.e. dimension--6 operators, results in the
following general form of the modification of any measured
observable $\mathcal{Z}$, in particular, cross section in terms of
the Wilson coefficients
\cite{JHEP-2002-131-2020,PU-65-653-2022,JHEP-2106-010-2021}
\begin{equation}
\displaystyle \mathcal{Z}^{\scriptsize{\mbox{EFT}}}=
\mathcal{Z}^{\scriptsize{\mbox{SM}}}+\biggl(\sum_{i}\frac{C_{i}^{(6)}}{\Lambda^{2}}\mathcal{Z}^{\scriptsize{\mbox{Int},(6)}}_{i}
+\mbox{h.c.}\biggr)+\biggl(\sum_{i,j}\frac{C_{i}^{(6)}C_{j}^{(6)}}{\Lambda^{4}}\mathcal{Z}^{\scriptsize{\mbox{BSM},(6)}}_{ij}+
\mbox{h.c.}\biggr),\label{eq:1.3add}
\end{equation}
where $\mathcal{Z}^{\scriptsize{\mbox{SM}}}$ is the SM prediction,
the second term contains the contributions
$\mathcal{Z}^{\scriptsize{\mbox{Int},(6)}}_{i}$ arising from the
interference of a single dimension--6 operator with the SM, the
third term arises from the interference of two diagrams containing
one dimension--6 operator each\footnote{The term
$\Lambda^{-4}$-order also appears as a consequence of the
interference of the dimension--8 operator contributions with the
SM ones, but this part does not included in (\ref{eq:1.3add})
because of truncation of $\mathcal{L}_{\scriptsize{\mbox{eff}}}$
established above.}, represents non-linear effects of new physics
only and, consequently, quantities
$\mathcal{Z}^{\scriptsize{\mbox{BSM},(6)}}_{ij}$ are due to purely
physics BSM.

As indicated above the dominant process is (\ref{eq:1.1.1}) for
the $s$--channel of single $t$ production through the EW
interaction. The LO partonic single $t$ production cross section
due to process (\ref{eq:1.1.1}) within EFT with leading
modification to SM process up to the $\Lambda^{-2}$-order terms is
\cite{PRD-83-034006-2011,JHEP-2002-131-2020}
\begin{eqnarray}
\displaystyle
&&\hat{\sigma}^{\scriptsize{\mbox{(0),EFT}}}_{u\bar{d} \to
t\bar{b}}(m_{W},m_{t},s)=
\hat{\sigma}^{\scriptsize{\mbox{(0),EW}}}_{u\bar{d} \to t\bar{b}}+
\hat{\sigma}^{\scriptsize{\mbox{(0),eff}}}_{u\bar{d} \to
t\bar{b}}=\\\nonumber &&= \Pi_{V}^{2}\frac{\pi
\alpha_{W}^{2}}{24s}\frac{\beta_{ts}^{4}(2+\rho_{ts})}{\beta_{Ws}^{4}}\biggl\{1+
\frac{2s\rho_{Ws}}{\Pi_{V} \pi
\alpha_{W}}\biggl(\frac{C_{\scriptsize{\mbox{$\phi$Q}}}^{(3)}}{\Lambda^{2}}+
\frac{\mbox{Re}C_{\scriptsize{\mbox{tW}}}}{\Lambda^{2}}\frac{3\sqrt{2\rho_{ts}}}{\sqrt{\rho_{Ws}}(2+\rho_{ts})}+
\frac{C_{\scriptsize{\mbox{Qq}}}^{(3,1)}}{\Lambda^{2}}\frac{\beta_{Ws}^{2}}{\rho_{Ws}}
\biggr)\biggr\},\label{eq:1.4}
\end{eqnarray}
where $C_{\scriptsize{\mbox{$\phi$Q}}}^{(3)}$,
$C_{\scriptsize{\mbox{tW}}}$ and
$C_{\scriptsize{\mbox{Qq}}}^{(3,1)}$ are the Wilson coefficients
for dimension--6 operators in
$\mathcal{L}_{\scriptsize{\mbox{eff}}}$.

According to the mostly used approach, the present work is focused
on CP--conserving extensions of the SM, assuming that all Wilson
coefficients are real and therefore neglecting CP--violating
interactions \cite{JHEP-2002-131-2020}. Based on the available
estimations, the following ranges are used for the Wilson
coefficients $C_{\scriptsize{\mbox{$\phi$Q}}}^{(3)} \in
[-1.145;0.740]$, $C_{\scriptsize{\mbox{tW}}} \in [-0.313;0.123]$,
$C_{\scriptsize{\mbox{Qq}}}^{(3,1)} \in [-0.163;0.296]$ in units
of $(\mbox{TeV}/\Lambda)^{2}$ in the present work and these ranges
correspond to the 95\% confidence level from the global
(marginalized) fit using linear in the $\Lambda^{-2}$ EFT
calculations \cite{JHEP-2111-089-2021} for self-consistency with
(\ref{eq:1.3add}) and (\ref{eq:1.4}). The corresponding median
values calculated as simple average of the boundary values are
$\langle C_{\scriptsize{\mbox{$\phi$Q}}}^{(3)}\rangle = -0.2 \pm
0.9$, $\langle C_{\scriptsize{\mbox{tW}}}\rangle = -0.10 \pm
0.22$, $\langle C_{\scriptsize{\mbox{Qq}}}^{(3,1)}\rangle = 0.07
\pm 0.23$ in units of $(\mbox{TeV}/\Lambda)^{2}$.

\section{Results}\label{sect:2}

The energy range for protons in laboratory reference system
considered in the present paper is $E_{p}=10^{17}$--$10^{21}$ eV.
This range includes the energy domain corresponded to the
Greisen--Zatsepin--Kuzmin (GZK) limit
\cite{Greisen-PRL-16-748-1966} and somewhat expands it, taking
into account, on the one hand, both possible uncertainties of
theoretical estimations for the limit values for UHECR and
experimental results, namely, measurements of several events with
$E_{p} > 10^{20}$ eV and the absence of UHECR particle flux
attenuation up to $E_{p} \sim 10^{20.5}$ eV
\cite{Okorokov-PAN-81-508-2018} and, on the other hand, the
energies corresponding to the nominal value
$\sqrt{\smash[b]{s_{\scriptsize{pp}}}}=14$ TeV of the commissioned
LHC as well as to the parameters for the main international
projects high energy LHC (HE--LHC) with the nominal value
$\sqrt{\smash[b]{s_{\scriptsize{pp}}}}=27$ TeV and Future Circular
Collider (FCC) with $\sqrt{\smash[b]{s_{\scriptsize{pp}}}}=100$
TeV. Therefore the estimations below can be useful for both the
UHECR physics and the collider experiments.

For sufficiently high collision energies at which
$\sqrt{\rho_{ts}} \ll 1$, the following limiting relations can be
used $\forall\,x=W,t: \rho_{xs} \to 0$, $\beta_{xs} \to 1$ and, as
consequence\footnote{The condition $\rho_{ts} \leq
\rho_{ts}^{\tiny{\mbox{max}}}$ can be used for estimating the
onset of the domain in which high-energy approach is valid, where
$\rho_{ts}^{\tiny{\mbox{max}}}$ is some empirical number. The
choice $\rho_{ts}^{\tiny{\mbox{max}}}=10^{-2}$ allows
$\sqrt{\smash[b]{s_{\tiny{\mbox{min}}}}}=10m_{t} \approx 1.72$ TeV
which corresponds to the
$\sqrt{\smash[b]{s_{\scriptsize{pp}}^{\tiny{\mbox{min}}}}} \approx
5.2$ TeV. Thus one can expect that the asymptotic relation
(\ref{eq:2.1}) will be valid for energy range under consideration
here.},
\begin{equation}
\displaystyle
\left.\hat{\sigma}^{\scriptsize{\mbox{(0),EFT}}}_{u\bar{d} \to
t\bar{b}}\right|_{\rho_{ts} \to 0} \longrightarrow
\Pi_{V}^{2}\frac{\pi \alpha_{W}^{2}}{12s}\biggl(1+\frac{2s}{\pi
\alpha_{W}\Pi_{V}}\frac{C_{\scriptsize{\mbox{Qq}}}^{(3,1)}}{\Lambda^{2}}\biggr).
\label{eq:2.1}
\end{equation}
Thus the relative contribution of leading modification to SM
process growths with the increase of  collision energy and can be
dominant for the LO partonic cross section for single $t$
production due to process (\ref{eq:1.1.1}) at finite value of the
Wilson coefficient in (\ref{eq:2.1}) at large $s \gg m_{t}^{2}$.

It should be emphasized the following important features of the
calculations within EFT. At present all available estimations
obtained from individual and global fits of experimental data have
a (very) large uncertainties and coincide with null within errors
for the Wilson coefficients in (\ref{eq:1.4}), (\ref{eq:2.1}). As
consequence the contribution of the leading modification to SM
process, strictly speaking, agrees with null for the single $t$
production in $s$--channel (\ref{eq:1.1.1}) at available accuracy
of measurements. The terms $\propto
C_{\scriptsize{\mbox{Qq}}}^{(3,1)}$ have an opposite signs for
$s$--channel (\ref{eq:1.1.1}) and for $t$--channels
(\ref{eq:1.1.2a},\ref{eq:1.1.2b}). Therefore it is expected that
the contributions of these channels will cancel or, at least,
decrease significantly the term $\propto
C_{\scriptsize{\mbox{Qq}}}^{(3,1)}$ in the total LO partonic
single $t$ production cross section within EFT with leading
modification to SM process. The quantitative verification is in
the progress for this qualitative expectation. Therefore the
consideration below will be focused on the cross section
estimations with median values of the Wilson coefficients taking
into account the above clarifications.

For numerical calculations all masses and CKM elements are from
\cite{PTEP-2022-083C01-2022}. Figure shows the energy dependence
of LO partonic cross sections within EFT for the channel
(\ref{eq:1.1.1}), where the dashed line corresponds to the
contribution from electroweak part of the SM (k=EW), the dotted
line -- to the term from the new physics effects (k=eff) and the
solid line is the sum for the single $t$ production in the
framework of the EFT (k=EFT). As emphasized above the curves for
k=eff and EFT are deduced for median value of
$C_{\scriptsize{\mbox{$\phi$Q}}}^{(3)}$,
$C_{\scriptsize{\mbox{tW}}}$ and
$C_{\scriptsize{\mbox{Qq}}}^{(3,1)}$. In the inner panel the LO
partonic cross sections are shown for the $s$--channel at energies
$\sqrt{\smash[b]{s_{\scriptsize{pp}}}}=1.8-14$ TeV which some
lower than the range under consideration for completeness
information. This range covers from the measurements at Tevatron
($\sqrt{\smash[b]{s_{\scriptsize{pp}}}}=1.8-1.96$ TeV) up to
nominal $\sqrt{\smash[b]{s_{\scriptsize{pp}}}}$ of the LHC and
adjoins to the energy domain under study. In accordance with
(\ref{eq:1.4}) the terms
$\hat{\sigma}^{\scriptsize{\mbox{(0),EW}}}_{u\bar{d} \to
t\bar{b}}$ and
$\hat{\sigma}^{\scriptsize{\mbox{(0),eff}}}_{u\bar{d} \to
t\bar{b}}$ show the opposite dependence on $s_{\scriptsize{pp}}$
in functional sense: the LO partonic cross section for channel
(\ref{eq:1.1.1}) decreases with the increase of the collision
energy as $\propto s_{\scriptsize{pp}}^{-1}$ at qualitative level
within SM whereas the contribution of the leading modification to
SM process increases with collision energy up to the
$\sqrt{\smash[b]{s_{\scriptsize{pp}}}} \simeq 10$ TeV and then it
is almost flat. The behavior of
$\hat{\sigma}^{\scriptsize{\mbox{(0),EFT}}}_{u\bar{d} \to
t\bar{b}}(s_{\scriptsize{pp}})$ at
$\sqrt{\smash[b]{s_{\scriptsize{pp}}}} \simeq 5-10$ TeV confirms
the qualitative estimation deduced above for the low boundary in
$s_{\scriptsize{pp}}$ of the domain of validity for the
high-energy approach (\ref{eq:2.1}). The SM mechanism gives the
dominate contribution in single $t$ production in $s$--channel in
the energy range from Tevatron up to the
$\sqrt{\smash[b]{s_{\scriptsize{pp}}}} \approx 3$ TeV and excess
of the contribution due to the physics BSM over SM growth at
higher energies rapidly. With taking into account the important
clarifications made above the cross section due to effective part
of the Lagrangian exceeds the corresponding parameter from SM
significantly for median values of the Wilson coefficients at
$\sqrt{\smash[b]{s_{\scriptsize{pp}}}}=14$ TeV already (Fig., main
panel). The flat behavior of
$\hat{\sigma}^{\scriptsize{\mbox{(0),EFT}}}_{u\bar{d} \to
t\bar{b}}(s_{\scriptsize{pp}})$ agrees with (\ref{eq:2.1}) and
implies validity of the high-energy asymptotic approach in full
main energy domain ($\sqrt{\smash[b]{s_{\scriptsize{pp}}}} \geq
13.7$ TeV) under study.

The uncertainty from uncalculated higher orders in the
perturbative expansion is estimated by varying the $\mu_{F,R}$
independently around the central scale choice, $\mu$. Usually, the
following intervals are considered for estimation of the
theoretical uncertainty \cite{JHEP-0910-042-2009}
\begin{equation}
\displaystyle (\mu_{F},\mu_{R}) \in
\bigl\{(\mu/2,\mu/2),(\mu/2,\mu),(\mu,\mu/2),(\mu,2\mu),(2\mu,\mu),(2\mu,2\mu)\bigr\}.\nonumber
\end{equation}
Also values of the SM--parameters $m_{W}$, $m_{t}$, $V_{ud}$,
$V_{tb}$ have errors and contribute to the uncertainties of the
cross sections $\Delta
\hat{\sigma}^{\scriptsize{\mbox{(0),k}}}_{u\bar{d} \to t\bar{b}}$
(k=EW, eff, EFT). Detailed analysis results in the conclusion that
the contributions of all these uncertainty sources are neglected
with respect to the errors due to (very) large uncertainties of
the Wilson coefficients and, consequently, $\Delta
\hat{\sigma}^{\scriptsize{\mbox{(0),EFT}}}_{u\bar{d} \to t\bar{b}}
\approx \Delta
\hat{\sigma}^{\scriptsize{\mbox{(0),eff}}}_{u\bar{d} \to
t\bar{b}}$. Because of relative errors for $\langle
C_{\scriptsize{\mbox{$\phi$Q}}}^{(3)}\rangle$, $\langle
C_{\scriptsize{\mbox{tW}}}\rangle$, $\langle
C_{\scriptsize{\mbox{Qq}}}^{(3,1)}\rangle$ larger than 1 there are
no uncertainty bands in Figure for clearness and only curves
obtained for median values of Wilson coefficients are shown as
described above.

\section{Conclusions}\label{sect:3}

Summarizing the foregoing, one can draw the following conclusions.

The partonic cross section for single $t$ production is considered
for the $s$--channel at LO level within both the SM and the EFT in
ultra-high energy range. The EFT approach takes into account the
dimension--6 operators.

The LO partonic cross sections differ significantly for SM and EFT
with median values of the Wilson coefficients for single $t$
production in the $s$--channel at nominal LHC energy already. The
LO partonic cross section for single $t$ production in the
$s$--channel within EFT is almost flat at the level on order 0.15
pb in the energy domain under consideration, i.e. up to the
highest energies $\mathcal{O}(1~\mbox{PeV})$.

The work is in the progress for other partonic processes
($t$--channel and $tW$ production) as well as for estimations of
hadronic cross sections for single $t$ production at ultra-high
energies.

\section*{Acknowledgments}

This work was supported in part within the National Research
Nuclear University MEPhI Program ''Priority 2030".

\newpage
\begin{figure*}
\includegraphics[width=16.0cm,height=16.0cm]{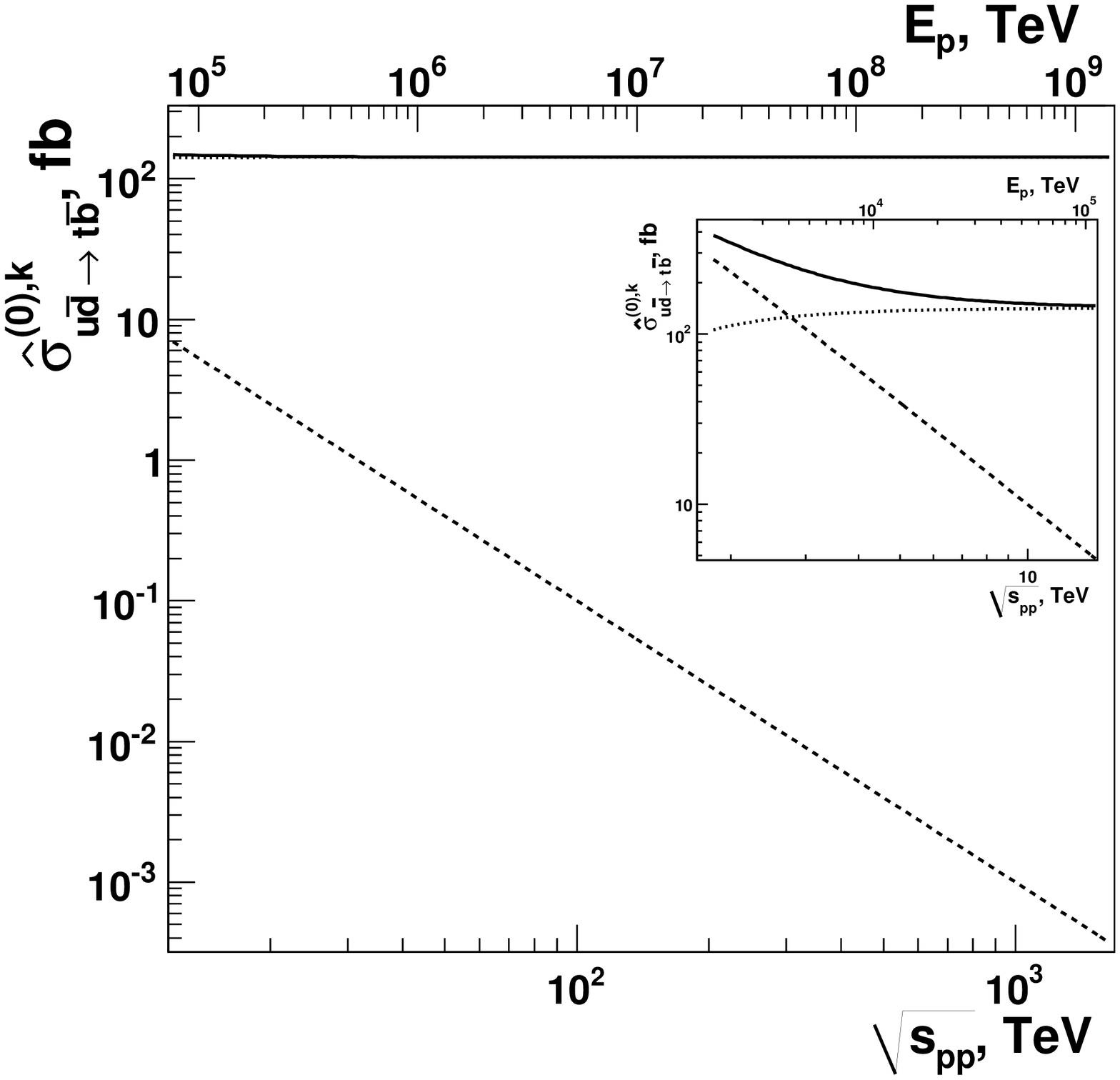}
\vspace*{8pt} \caption{Energy dependence of LO partonic cross
sections within EFT for the channel (\ref{eq:1.1.1}). The dashed
line corresponds to the contribution from electroweak part of the
SM (k=EW), the dotted line -- to the term from the new physics
effects (k=eff) and the solid line is the sum for the single $t$
production in the framework of the EFT (k=EFT). Curves for k=eff
and EFT are deduced for median value of
$C_{\scriptsize{\mbox{$\phi$Q}}}^{(3)}$,
$C_{\scriptsize{\mbox{tW}}}$ and
$C_{\scriptsize{\mbox{Qq}}}^{(3,1)}$. Inner panel: LO partonic
cross sections for the $s$--channel at lower energies
$\sqrt{\smash[b]{s_{\scriptsize{pp}}}}=1.8-14$ TeV.} \label{fig:1}
\end{figure*}

\end{document}